\documentstyle[12pt,epsf]{article}

\date{\today}

\newcommand\putfig[3]{
   \vbox{
   \let\picnaturalsize=N
   \def\picsize{#3}
   \def\picfilename{#1}
   \ifx\nopictures Y\else{\ifx\epsfloaded Y\else\input epsf \fi
   \let\epsfloaded=Y
   \centerline{\ifx\picnaturalsize N\epsfxsize \picsize\fi
   \epsfbox{\picfilename}}}\fi
   \vspace{1.0cm}
   {\it #2}
   \vspace{1.5cm}
   }
}

\def\be{\begin{equation}}
\def\ee{\end{equation}}
\def\bear{\begin{eqnarray}}
\def\eear{\end{eqnarray}}
\def\nn{\nonumber}

\def\bra{{\langle}}
\def\ket{{\rangle}}
\def\hlf{{{1\over 2}}}

\def\lbr{{\lbrack}}
\def\rbr{{\rbrack}}

\def\wdg{{\wedge}}                              

\newcommand\px[1]{{\partial_{#1}}}
\newcommand\qx[1]{{\partial^{#1}}}

\newcommand\inv[1]{{1\over{#1}}}

\newcommand\rep[1]{{\underline{\bf {#1}}}}      
\newcommand\tr[1]{{\mbox{tr}\{{#1}\}}}          
\newcommand\ev[1]{{\bra {#1} \ket}}             
\newcommand\evtr[1]{{\bra \tr{{#1}} \ket}}      
\newcommand\com[2]{{\lbrack {#1},{#2}\rbrack}}  

\def\IZ{{\bf Z}}                                

\newcommand\MR[1]{{{\bf R}^{#1}}}               
\newcommand\MC[1]{{{\bf C}^{#1}}}               
\newcommand\MS[1]{{{\bf S}^{#1}}}               
\newcommand\MT[1]{{{\bf T}^{#1}}}               

\def\a{{\alpha}}
\def\b{{\beta}}
\def\g{{\gamma}}
\def\d{{\delta}}
\def\u{{\mu}}
\def\v{{\nu}}

\def\s{{\sigma}}
\def\t{{\tau}}
\def\h{{\eta}}

\def\lam{{\lambda}}

\newcommand\dlu[1]{{\delta^{#1}}}               

\newcommand\ep[1]{{\epsilon_{#1}}}              

\newcommand\lylx[2]{{{\delta {#1}} \over {\delta {#2}}}} 
\newcommand\llylxx[3]{{{\delta^2 {#1}} \over {\delta {#2} \delta {#3}}}}

\def\suv{{\sigma_{\u\v}}}                       

\newcommand\path[1]{{\Gamma_{#1}}}              

\newcommand\wil[1]{{W\lbr {#1} \rbr}}           

\newcommand\surf[1]{{{\cal W}\lbr {#1} \rbr}}   
\newcommand\osurf[1]{{\widehat{\cal W}\{ {#1} \}}} 
\newcommand\osurfi[2]{{\widehat{\cal W}_{#1}\{ {#2} \}}}

\newcommand\SUSY[1]{{{\cal N}= {#1}}}           

\newcommand\lnk[1]{{{\cal L}({#1})}}            

\def\Ham{{\widehat{\cal H}}}                    
\def\npb#1#2#3{{\it Nucl.\ Phys.} {\bf B#1} (19#2) #3}
\def\plb#1#2#3{{\it Phys.\ Lett.} {\bf B#1} (19#2) #3}

\def\prp#1#2#3{{\it Phys.\ Rep.} {\bf #1} (19#2) #3}

\def\hepth#1{{\it hep-th/{#1}}}

\begin{document}
\begin{titlepage}
\titlepage
\rightline{PUPT-1610}
\rightline{hep-th/9605201}
\rightline{May 27, 1996}
\vskip 1cm
\centerline {{\Large \bf Six-Dimensional Tensionless Strings}}
\centerline {{\Large \bf In The Large $N$ Limit}}
\vskip 1cm
\centerline {Ori J. Ganor}
\vskip 0.5cm
\begin{center}
\em  origa@puhep1.princeton.edu\\
Department of Physics, Jadwin Hall, Princeton University\\
Princeton, NJ 08544, U. S. A.
\end{center}
\vskip 1cm
\abstract{
 When $N$ five-branes of M-theory coincide
the world-volume theory contains tensionless strings,
according to Strominger's construction.
This suggests a large $N$ limit of tensionless string theories.
 For the small $E_8$ instanton theories, the definition would
be a large instanton number.
An adiabatic argument suggests that in the large $N$ limit 
an effective extra uncompactified dimension might be observed.
We also propose ``surface-equations'',
which are an analog of Makeenko-Migdal loop-equations,
 and might describe correlators in the tensionless string theories.
In these equations, the anti-self-dual two forms of 6D and 
the tensionless strings enter on an equal footing.
Addition of strings with CFTs on their world-sheet 
is analogous to addition of matter in 4D QCD.
}
\end{titlepage}

\section{Introduction}

Although all known quantum field theories in 6D are trivial in
the IR limit, it has been suggested \cite{WitCOM}, on the
basis of non-perturbative string theory, that non-trivial
quantum theories in flat $\MR{5,1}$ do exist.
These theories have been called {\em ``tensionless non-critical
strings''}.
They acquired that name because by 
a relevant perturbation (Higgs-ing the tensor multiplet)
of the tensionless non-critical string theory we can reach
the ``tensile non-critical string''. That theory is described
in the IR by a free anti-self-dual tensor field.
The massive BPS spectrum contains a 1-brane which is a source
for the 2-form field.

There are various constructions of such theories from string
theory.

In \cite{WitCOM} it was argued that when type-IIB is compactified
on a $K_3$ with a  very small 2-cycle the resulting 6D theory
contains a BPS 1-brane with a tension proportional to the area
of the 2-cycle. When the 2-cycle shrinks to zero and the $K_3$
develops an $A_1$ singularity the string becomes of zero tension.
In \cite{StrOPN} the same theory was constructed as the
world volume theory of two parallel 5-branes in M-theory that 
become very close. The connection between the two examples
was made explicit in \cite{WitFIV}.
This example has $\SUSY{2}$ SUSY in 6D.

An example with $\SUSY{1}$
was constructed in \cite{GanHan} as a small heterotic 
$E_8\times E_8$ instanton and an extensive description
of $\SUSY{1}$ tensionless string theories 
that arise at phase transition points of heterotic vacua
on $K_3$ was given in \cite{SWCSD}.

All the constructions as well as new ones
can be described in the framework of 
 F-theory \cite{VafaFT,SWCSD,WitPMF,MVI,MVII} as 
type-IIB vacua on a base manifold with a 2-cycle
that shrinks to zero.

What are the variables in terms of which six-dimensional
non-critical string theories should be formulated?

It is unknown if a Lagrangian formulation exists.

Could there be a world-sheet formulation as a string
theory?
In that case there are some features of the theories that
seem very hard to incorporate in a world-sheet description:
\begin{enumerate}
\item
The graviton does not appear in the spectrum \cite{WitCOM}.
\item
The non-critical string is charged under an anti-symmetric
anti-self-dual tensor field $B_{\u\v}^{(-)}$ \cite{DFKR}.
It is thus dual to itself in the ``electric-magnetic duality'' 
sense \cite{WitCOM}.
\item
The string coupling constant is fixed and is inevitably 
not weakly coupled. 
This follows from the anti-self-duality of the string
and the Dirac quantization condition \cite{WitCOM}.
\item
The tension of the string 
can be taken to be either non-zero
or zero according to the VEVs of the tensor multiplet.
When the tension vanishes we obtain a scale invariant theory
that is non-trivial \cite{WitCOM,SWCSD}.
\item
Compactifying to 4D on a torus there are no multiply-wound modes
of the string (except $(p,q)$ for co-prime $p$ and $q$
which is a winding number 1 on another 1-cycle) \cite{KlePRV}.
\end{enumerate}
In particular, the fact that in 6D the genus expansion
is not an expansion in a small parameter seems to suggest
that any world-sheet approximation would be outside
its region of applicability.

One of the motivations to this paper was to explore a third possibility
which is neither a field-theory nor a world-sheet theory.

We know that upon compactification to 4D on a torus $\MT{2}$
one obtains a gauge theory \cite{WitCOM}. 
As argued in \cite{StrOPN},
when $N$ 5-branes of the M-theory come close together there emerges
a six-dimensional world-volume theory that contains $N$
tensor multiplets and $N(N-1)$ kinds of strings on the 6D world-volume
which correspond to 2-branes connecting the 5-branes.
When this theory is compactified to 4D one obtains 
$U(N)$ $\SUSY{4}$ Yang-Mills.
The gauge group is generically broken to $U(1)^{N}$.
The $N$ photon fields come from the dimensional reduction
of the $B_{\u\v}^{(-)}$ that live on the 5-branes.
The $(N^2-N)$ W-bosons in 4D arise as winding modes of the 6D strings
around one of the cycles of $\MT{2}$. Winding  states around
other cycles will give monopoles and dyons.

When the $N$ 5-branes coincide we obtain the scale invariant
theory in 6D that gives rise to the un-Higgsed $\SUSY{4}$ $U(N)$ in 4D.

It is interesting that in 4D the winding modes of the six-dimensional
tensionless strings as well as the reduction of the 2-forms
are related by a non-abelian gauge symmetry.
In an $SU(2)$ gauge theory there is no real distinction between
the $W^3$ boson and $W^{\pm}$. This suggests that maybe in 6D
there is a formulation where the 2-forms $B_{\u\v}^{(-)}$
and the tensionless strings enter {\em on an equal footing}.

Gauge theories can be formulated in terms of geometrical variables
namely the Wilson loops. The variables of the theory are labeled
by loops and the correlators are related by the Makeenko-Migdal
loop equations \cite{MakMig}. Furthermore, in the large $N$ limit the
whole formulation simplifies because relations among loops (that
say wind $N$ times) that are 
a consequence  of the finite $N$ nature of the matrices
might be neglected. A further simplification arises because different
closed loops are uncorrelated up to $O(\inv{N^2})$.

It is quite possible that some of the features of tensionless
string theories simplify in the large $N$ limit.
We will argue, on the basis of adiabatic considerations
for type-IIB on an $A_{N-1}$ singularity
that the 6D tensionless string theory effectively
grows an extra dimension.
An adiabatic argument
that is similar in spirit
can be given for the small $E_8$ instanton string as well,
by using the CHS solution \cite{CHS} at a large instanton number.

We also wish to propose surface equations that maybe describe
correlators of degrees of freedom of the tensionless strings
in the limit when $N\rightarrow\infty$.
The surfaces would be non-abelian generalizations of 
the integrals $e^{i\int_S B_{\u\v}^{(-)}}$.

Our guidelines in guessing the equations were the requirements
of locality and unitarity.

The anti-self-duality of $B_{\u\v}^{(-)}$ causes extra complications
because the time derivative is first order and the Bianchi identity
is no longer satisfied.
We will discuss first an operator algebra that leads to surface
equations which are ``interacting'' generalizations of 
the equations that a free 2-form field should satisfy.
Then we will discuss the more complicated case of 
anti-self-duality.

The paper is organized as follows:
\begin{itemize}
\item
Section (2): A review of the constructions of
             tensionless string theories from non-perturbative
             string theory.
\item
Section (3): A generalization of loop equations
             to surface equations.
\item
Section (4): Anti-self-duality is incorporated  into the equations.
\item
Section (5): We add a few words about the super-symmetric case.
\item
Section (6): An adiabatic argument based on the CHS solution at 
             instanton number $N$ and
             a similar argument for type-IIB on an $A_{N-1}$ singularity,
             both for large $N$, is presented. We argue that an extra
             effective dimension might arise.
\item
Section (7): Discussion.
\end{itemize}

Section (6), is independent of the previous sections,
but we decided to postpone it till after the description
of the surface equations  in order to discuss how
the implications of an extra dimension can be
visible in the surface correlators.

I recently learned of another work that is under 
preparation \cite{ArgDie}, which discusses various aspects
of possible world-sheet formulations of tensionless strings
and might have some overlap with this introduction.

\section{Review of tensionless strings}

The existence of 
a tensionless string theory was first deduced in \cite{WitCOM}
by considering compactification of type-IIB on a $K_3$ 
with a 2-cycle whose area is very small compared to the
string scale. The resulting spectrum in 6D contains an $\SUSY{2}$
tensor multiplet with an anti-self-dual 2-form $B_{\u\v}^{(-)}$.
It is obtained from
that mode of the 4-form of type-IIB which is proportional to
the harmonic anti-self-dual 2-form in the $K_3$ Poincar\'e
dual to the small 2-cycle. The spectrum also contains a BPS
1-brane obtained by wrapping the 3-brane of type-IIB around
the 2-cycle.
 This 1-brane is charged under the $B_{\u\v}^{(-)}$.
When the area of the 2-cycle is much smaller than the string
scale the 2-form $B_{\u\v}^{(-)}$ (and its super-partner)
and the BPS 1-brane, which has a very small tension, can be
considered separately from the other string modes
for energies much smaller than the string scale. 
It was thus predicted that there exists a 6D theory which describes
a string interacting with an anti-self-dual 2-form but
which does not have the graviton in its spectrum.
Such a string was previously constructed in \cite{DFKR} as
a solitonic object.

Another construction of the same theory appeared in \cite{StrOPN}.
There it was shown that a 2-brane of M-theory can end
on the 5-brane. When $N$ 5-branes are very close an IR observer
sees a 6D theory which contains $N$ tensor multiplets
and $(N^2-N)$ strings. Each string is related to a 
2-brane that connects two of the $N$
5-branes (the strings have an orientation). The tension of the
string is proportional to the distance between the 5-branes
that it connects.
It was further shown in \cite{StrOPN,DougBB} that the endpoint
of a 2-brane is a source for $B_{\u\v}^{(-)}$ on the 
corresponding 5-brane. Thus each string in the 6D theory is
charged under the difference between the two $B_{\u\v}^{(-)}$
corresponding to the two 5-branes that it connects.

The two constructions -- as the world-volume theory on the
5-brane and type-IIB on $K_3$ -- were shown to be equivalent
in \cite{WitFIV} as a consequence of the equivalence
of M-theory on $T^5/Z_2$ with 16 
5-branes and type-IIB on $K_3$ \cite{DasMuk,WitFIV}.
When two 5-branes approach each other on the M-theory side,
there is a 2-cycle that shrinks in the $K_3$ on the 
type-IIB side.

In these constructions the string has a small tension.
When the $K_3$ is singular with an $A_{N-1}$ singularity or
when the $N$ 5-branes coincide, the string in 6D
becomes ``figuratively'' tensionless.
A more precise statement would be that the 6D theory 
has non-trivial IR dynamics -- a property that no known
quantum field theory has in 6D.

The above construction has $\SUSY{2}$ in 6D and upon
compactification on a torus $\MT{2}$ and going to the
IR limit, one obtains the non-trivial scale invariant
$\SUSY{4}$ Yang-Mills theory in 4D.
The modular parameter $\tau$ of the torus becomes the
coupling constant and $\theta$-angle of Yang-Mills
and it was suggested in \cite{WitCOM} that this would
be a natural explanation for S-duality of $\SUSY{4}$
Yang-Mills.

When the scalars of $\SUSY{4}$ YM get Higgsed,
the gauge group can be broken  to a maximal abelian 
subgroup with massive charged particles (the $W$-bosons)
and massive monopoles.
This theory would be obtained by compactifying
the 6D theory of strings with small tension on a torus
$\MT{2}$ of size much smaller compared to the tension.
The $W$-bosons and monopoles would correspond to 
strings in 6D winding around the different cycles
of $\MT{2}$ \cite{WitCOM}.

So far the constructions had $\SUSY{2}$ in 6D.
A tensionless string theory with less supersymmetry, 
namely $\SUSY{1}$ can be constructed from a heterotic 
$E_8$ instanton of zero size \cite{GanHan}.
Upon compactification of this theory to 4D one obtains
$\SUSY{2}$ Yang-Mills with matter. The gauge group is
$Sp(k)$ and there is some  matter in the 
fundamental $\rep{2k}$ of $Sp(k)$ (as in \cite{WitSML}).
In 6D there are additional tensionless
strings with a CFT on their world-sheet and those
are the ones that go over to the matter in 4D
\cite{GanHan,SWCSD,WitPMF}.

A unifying framework for all the above constructions
was given by F-theory \cite{VafaFT}.
Tensionless string theories arise when F-theory
is compactified on a base manifold with a shrinking 2-cycle.
The $E_8$ tensionless string arises on a 4D manifold
with a blown-up point \cite{SWCSD,WitPMF,MVI},
and there are  other constructions as well\cite{WitPMF,MVII}.

Tensionless string theories can be phase-transition
points in the moduli space of $\SUSY{1}$ theories in 6D
as was shown in \cite{SWCSD}.
In the example of a small $E_8$ instanton, the tensionless
string theory connects the ``nonzero-instanton'' phase
with another  phase that contains 1 more tensor multiplet
and 29 less hyper-multiplets.

Understanding the spectrum of states of tensionless string theory
seems to be also relevant for a microscopic understanding of 
the entropy of certain cases of black holes \cite{KleTse}.
It has also been suggested recently that a six-dimensional non-critical
string theory (though not tensionless) can describe the world-volume
theory of the 5-brane of the M-theory \cite{VVDBPS}.

To summarize, we repeat again some of the properties
of the non-critical string theories:
\begin{enumerate}
\item
The graviton does not appear in the spectrum of the 
non-critical string \cite{WitCOM}.
\item
The non-critical string is charged under an anti-symmetric
anti-self-dual tensor field $B_{\u\v}^{(-)}$ \cite{DFKR}.
It is thus dual to itself in the ``electric-magnetic duality'' 
sense \cite{WitCOM}.
\item
The string coupling constant is fixed and is inevitably 
not weakly coupled. 
This follows from the anti-self-duality of the string
and the Dirac quantization condition \cite{WitCOM}.
\item
The tension of the string can be taken to be either non-zero
or zero. 
When the tension vanishes we obtain a scale invariant theory
that is non-trivial \cite{WitCOM,SWCSD}.
\item
The theory can have either $(1,0)$ or $(2,0)$ super-symmetry
in six-dimensions.\footnote{I do not know if no super-symmetry 
at all is possible or is ruled out.}
When the super-symmetry is $(2,0)$ (which reduces to $N=4$ in 4D)
the possibilities seem to be limited \cite{WitCOM,StrOPN}.
On the other hand, there might be a diversity of $(1,0)$ theories.
In particular, those theories can carry a world-sheet current 
algebra \cite{GanHan,SWCSD,WitPMF}.
\item
Non-critical string theories in 6D provide insight into 4D
gauge theories and their BPS states \cite{WitCOM,KLMVW}.
\item
Tensionless string theories can be phase transition points
between two phases in 6D $\SUSY{1}$ theories \cite{SWCSD}.
\end{enumerate}

\section{Higher dimensional generalizations of Yang-Mills theory}

Yang-Mills  theories lead to a set of loop equations \cite{MigREV}.
Although it is hard to specify the exact boundary conditions
of these equations, they provide a geometrical realization
of the theory in terms of a complete set of gauge invariant 
quantities which are the Wilson loop expectation values:
\be
\wil{C} = \evtr{P e^{i\oint_C A_\u dx_\u}}.
\ee
and
\be
\wil{C_1,C_2,\dots} = 
\ev{\tr{P e^{i\oint_{C_1} A_\u dx_\u}}
    \tr{P e^{i\oint_{C_2} A_\u dx_\u}}
    \cdots}
\ee
The equations for $SU(N)$ are 
(see \cite{MigREV} for a comprehensive review):
\be
\px{\u}(x)\lylx{}{\s_{\u\v}(x)}\wil{C}
= g_B^2 N\oint_C dy_\v \dlu{(D)}(x-y) \{\wil{C_{xy}, C_{yx}}
  -\inv{N}\wil{C}\}
\label{YMleq}
\ee
$\lylx{}{\s_{\u\v}(x)}$ is the ``area-derivative''
and $\px{\u}(x)$ is the ``path-derivative''.

The non-abelian nature of Yang-Mills theories is captured
by a non-linear term in the loop equations which describes
the splitting of a self-intersecting loop as in Fig 1.
\vskip 0.5cm
\putfig{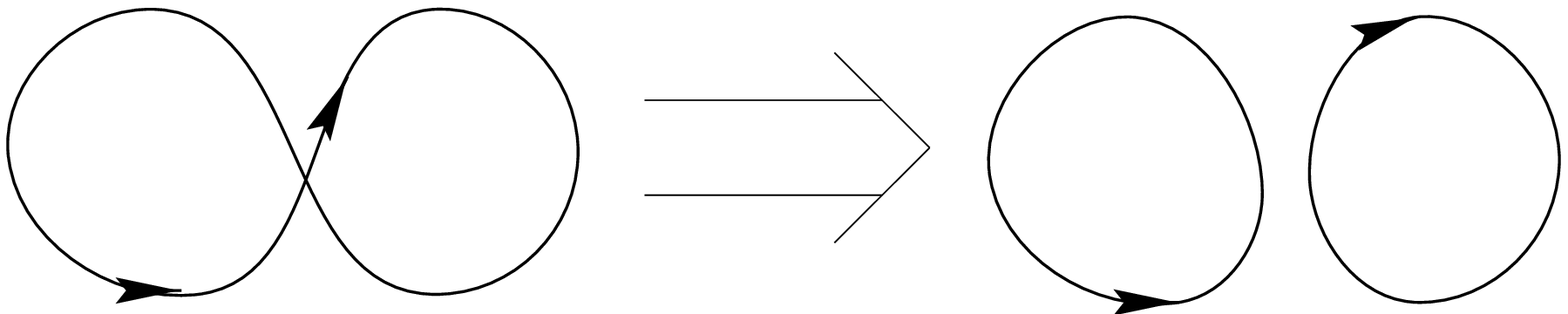}{\it Fig.1: The splitting of a self-intersecting
     loop in the RHS of the loop equations.}{100mm}
\vskip 0.5cm

Naive scaling dimensions of (\ref{YMleq}) imply that the 
coupling constant $g_B$ is dimensionless in 4D.
In higher dimensions Yang-Mills theories are non-renormalizable
and are IR trivial. 

However, we can contemplate a different kind of generalization
of (\ref{YMleq}) from 4D to higher dimensions.
Instead of using correlators that are parameterized
by loops we can try to use correlators that are labeled by
surfaces in $D$ dimensions. The non-linear term on the RHS
can be straightforwardly generalized to surfaces and we obtain
a set of equations with a naive scaling dimension that
makes the coupling constant dimensionless in 6D.

The resulting surface equations will be described below,
but we wish to precede them with a discussion of
their relevance to tensionless strings.
We will begin with what we believe is a direct
but {\em inappropriate} approach to tensionless strings
coupled to an anti-symmetric 2-form.
In this direct approach the strings and the
2-form do not enter on an equal footing.
In the subsection that will follow we will  remedy
that asymmetry by introducing the non-linear surface
equations.

Throughout this chapter we will simplify our lives
by leaving outside both super-symmetry and the anti-self-duality
of the 2-form. That is, the ``abelian part''  of the equations
will correspond to a full two-form with both self-dual
and anti-self-dual parts, and hence a Lagrangian description.

The next section will be devoted to the anti-self-dual theories.

\subsection{Interacting tensionless strings and two-forms}

\subsubsection{Abelian surface equations}
Let us start with a free field theory of 2-forms in 6D.
The field $B_{\u\v}$ is an anti-symmetric tensor with
action
\be
S = \hlf\int |dB|^2 d^6x
\ee
This action is gauge invariant under
\be
B_{\u\v} \rightarrow B_{\u\v} + \px{\u}\Lambda_\v - \px{\v}\Lambda_\u.
\ee
Instead of Wilson loops in 4D we have ``Wilson-surfaces'':
\be
\surf{S} = e^{i\int_S B}
\ee
The surface equations are derived in the usual way \cite{MigREV}:
\bear
\px{\a}\lylx{}{\s_{\a\u\v}(x)}\surf{S} &=&
\gamma_{\u\v}(S) \,\surf{S}    \label{Abel6D:e}\\
\ep{\a\b\g\d\u\v} \px{\a}\lylx{}{\s_{\b\g\d}(x)}\surf{S} &=& 0 
\label{Abel6D:m}
\eear
where
\be
\gamma_{\u\v}(S) dx^\u\wdg dx^\v
\ee
is a 2-form that is supported on the surface $S$ and is Poincar\'e
dual to the surface.
The first equation (\ref{Abel6D:e}) is derived from the equation
of motion
\be
d^{*}\,dB = 0
\ee
while the second is the Bianchi identity $dK=0$ where $K=dB$.

\subsubsection{Stringy matter in 6D}
Incorporation of dynamical matter fields in 4D QED is accomplished
by adding correlators of open loops. The end-points
are thought of  as creation and annihilation operators for
charged particles.
 For scalar particles, each end is supplemented with the equation
\be
\px{\u}(x)\qx{\u}(x)\ev{\Phi(C_{xy})} = m^2 \ev{\Phi(C_{xy})}
\ee
Here $\px{\u}$ is the {\em path derivative} defined in \cite{MigREV}
and $m$ is the mass of the particle at the corresponding end.

The equation for an ``electric'' $B_{\u\v}$ source is (formally):
\be
\px{\a}\lylx{}{\s_{\a\u\v}(x)}\surf{S} = 
\surf{S,\Theta^{(e)}_{\u\v}(x)}
+\gamma_{\u\v}(S) \,\surf{S}
\ee
$\Theta^{(e)}_{\u\v}(x)$ is the ``electric'' current.

It would seem that the six-dimensional generalization of QED with
matter would be to include both closed and open surfaces
in the correlators. The boundaries of open surfaces will 
describe creation and annihilation of ``stringy'' matter.

In the ``particle'' case in 4D, there was
an equation in terms of derivatives of the boundary position
(i.e. the end-points of the loops) which expressed 
the mass-shell condition $p^2 = m^2$ for the end-point particles
(in the scalar case). What is the analog of that term in 6D for
strings? 

The operator that is the 6D analog of $p^2$
should be given by a second-order
differential operator on the boundary and as such has to 
be constructed from the {\em loop} differential operators
\be
\lylx{}{\suv(x)},\,\,\,\, \px{\u}(x)
\ee
The RHS should be the tension $T$ of the string.

We can try to make the following  Ansatz.
Let us consider a specific boundary $C$ of the surface $S$.
$C$ is a simple closed loop. Forgetting for a moment about $S$,
let $C$ be parameterized by $x^\u(s)$. Let us fix
the reparameterization invariance by choosing $s$ to be the
length of the string.
A piece of the string from $s$ to $s+ds$ has mass $T ds$.
On the other hand, its momenta in the directions transverse to
the string are 
\be
P_\u = (ds)\times ({{dx^\v}\over {ds}})\lylx{}{\suv(x)}
\ee
So we can implement
\be
T^2 \h_{\u\v} dx^\u dx^\v = T^2 ds^2 = \h_{\a\b}P^\a P^\b = 
\h_{\a\b} dx^\v dx^\u \llylxx{}{\s_{\u\a}(x)}{\s_{\v\b}(x)}
\ee
we find formally:
\be
\h_{\a\b}\llylxx{}{\s_{\u\a}(x)}{\s_{\v\b}(x)}\surf{S}
= \h_{\u\v} T^2 \surf{S}
\ee
where the differentiations act on the boundary.
There is however the issue of divergences. We know from \cite{MigREV},
for example, that two area derivatives at the same point is an
ill-defined object.
However, it might be that a renormalization of the tension $T$
will be sufficient.
\footnote{I am grateful to David Gross
          for suggesting this possibility.}

Another possibility is that with enough super-symmetry there
will not be a divergence.

\subsubsection{Analogy with Yang-Mills theories}
Let us re-assess the situation so far by recalling the
guideline -- the tensionless string theory in 6D.

The ``non-critical string''
has twice the minimum $B_{\u\v}^{(-)}$ charge. It is analogous
to the $W^\pm$ gauge bosons. Indeed after compactification
to 4D the winding states give rise to $W^\pm$ and $B_{\u\v}^{(-)}$
will give rise to $W^3$.
It follows that at  each string boundary {\em two}  (overlapping)
sheets must end.
In the discussion up to now, the surface should have really
been interpreted as {\em double layered}.
 But now we can just as well allow two different single layer
sheets to end at each boundary. Allowing them to be
non-overlapping we end up with configurations with
surfaces on which there are strings (the ``strings''
are what we called up to now ``boundaries'').
On a surface, each string
is the boundary of two sheets (inside and outside) which agrees
with the fact that the string has twice the minimum $B_{\u\v}^{(-)}$
charge.

However,
if we would have applied a ``similar'' reasoning for 4D Yang-Mills
we would have ended up with a $U(1)$ gauge theory coupled to 
the $W^\pm$ ``matter''. Instead of closed non-abelian Wilson loops
we would have used closed $U(1)$ Wilson loops and open loops
ending on $W^\pm$ points. We know however, that this is not
the correct formulation of $SU(2)$ Yang-Mills. 
The correct formulation has just closed loops and the $W^\pm$
and $W^3$ are treated on an {\em equal footing}.

We are lead to believe that we have done the same error in 6D.
It would thus seem that the correct set of variables in 6D
is just the set of {\em closed} surfaces. The ``marked'' strings 
that we drew on them before
``mingle'' with the $B_{\u\v}^{(-)}$ fields and together they form
one surface operator, just like the $W^\pm$
bosons and $W^3$ enter on an equal footing in a Wilson loop in 4D.

\subsection{Surface equations}
In the large $N$ limit, the variables will be functionals $\surf{S}$
of {\em surfaces} $S$ embedded in 6D. The surfaces are oriented but can
be of arbitrary genus.
The surfaces are defined by a map from some ``world-sheet'' Riemann
surface of genus $g$ to $\MR{5,1}$ with the equivalence of surfaces
that differ just by reparametrization.
This ``world-sheet'' is related to the surface-variable and is {\em not}
the world-sheet of the tensionless string.
 For the moment we will assume that there are no anomalies 
with respect to reparameterization.
We think this is plausible as long as we don't deal with
a chiral theory.
(We will return to the question of gravitational anomalies on the
world-sheet later  for the anti-self-dual case. Indeed, in that
case we expect gravitational anomalies to be canceled only after
addition of tensionless strings with a CFT with $c-\bar{c}=8$
on their world-sheet,
though we were not able to prove it in this framework.)

Surfaces can self-intersect.
because we used a world-sheet in the definition,
two surfaces that look the same as sets in 6D might be inequivalent
(and this can happen only if the surfaces self-intersect).

Similarly to large $N$ 4D QCD Wilson loop rules,
we might expect that a functional of a surface that is disconnected is,
up to $O(\inv{N^2})$ corrections, the product
of the functionals of its connected components.

There are only {\em closed} surfaces.

A direct generalization of the area-derivative $\lylx{}{\suv}$
for loop functionals is the {\em volume-derivative} for surface
functionals:
\be
\lylx{}{\s_{\u\v\t}(x)}\surf{S}
\ee
which is defined by adding to the surface a small bubble in
the direction $\u\v\t$ at the position $x$ on the surface,
and then calculating the change in $\surf{S}$ and dividing by
the volume of the bubble.

We also need to define an operator $K_p(S)$ which acts on surfaces
$S$ where $p$ is a point of self-intersection of the surface.
$K_p(S)$ will be another surface
that {\em as a set in 6D} is identical to $S$, but has a 
different world-sheet map and thus is distinct from $S$ in our sense.
It is defined as follows. Let us consider the vicinity
of the self-intersection point $p$. The surface will look like
two planes that intersect transversely. Let us denote the planes
by $x_1-x_2$ and $x_3-x_4$ (see Fig 2).
\vskip 0.5cm
\putfig{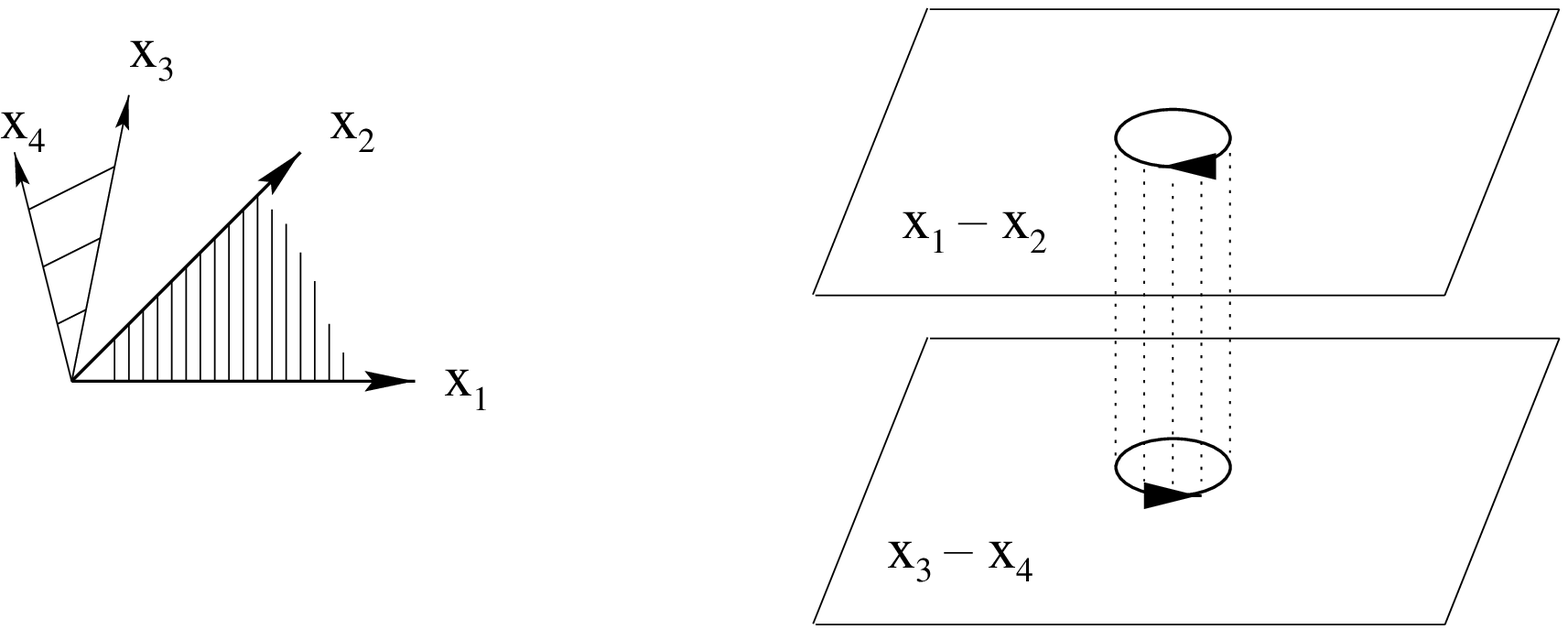}{\it Fig.2: The transverse intersection
                      of two surfaces}{100mm}
\vskip 0.5cm
The intersection point will be the origin
in each of the planes. Now cut a small disk of radius $\epsilon$
around each of the origins of the two planes and glue the two planes
along the circles of radius $\epsilon$ keeping track of orientation.
As we take $\epsilon\rightarrow 0$ we obtain $K_p(S)$.
The resulting surface is described in Fig 3.
\vskip 0.5cm
\putfig{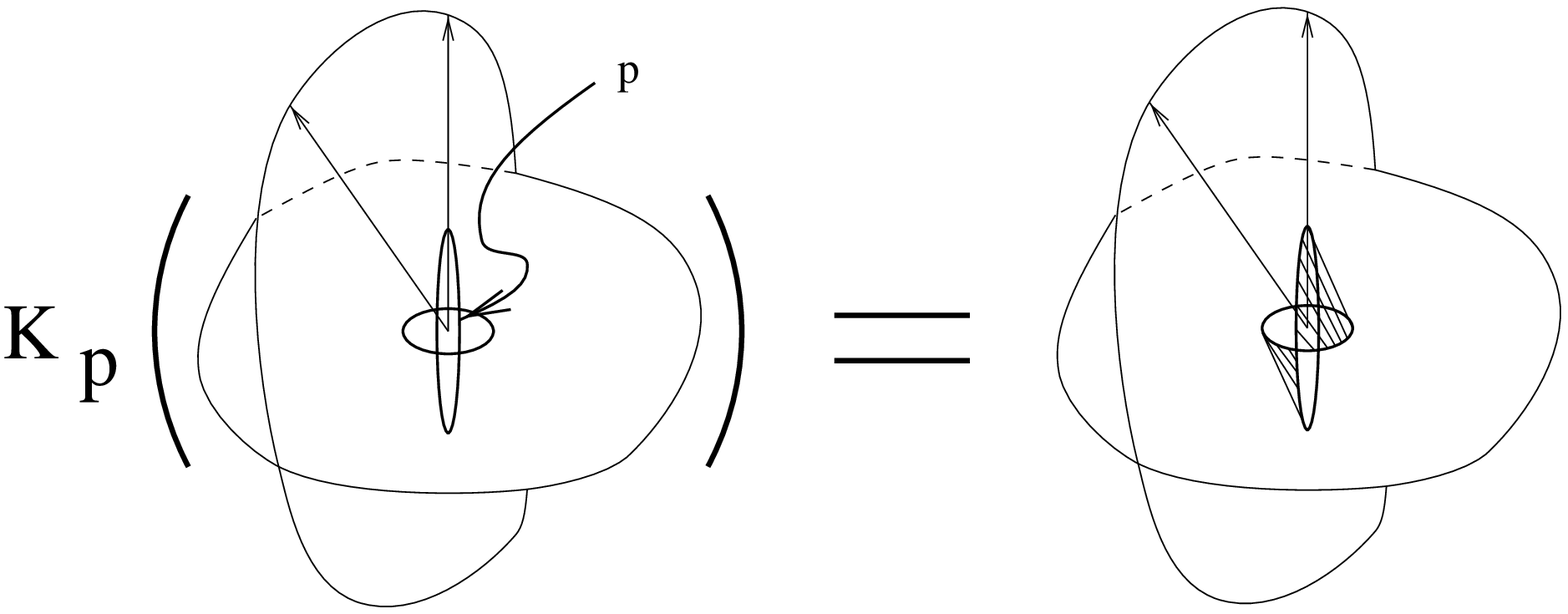}{\it Fig.3: Graphical description of 
                         the $K_p$ operator}{100mm}
\vskip 0.5cm
This procedure is a direct generalization of the way the RHS of
the QCD loop equation is defined.

Heuristically we might interpret the construction of $K_p$
as an infinitesimal string that is exchanged between
the two surfaces at the point of intersection.

The $K_p$ operator increases the genus of $S$ by one.
 For example,  it transforms a sphere that is smashed
such that its north pole and south pole touch -- into 
a torus (Fig 4).
\vskip 0.5cm
\putfig{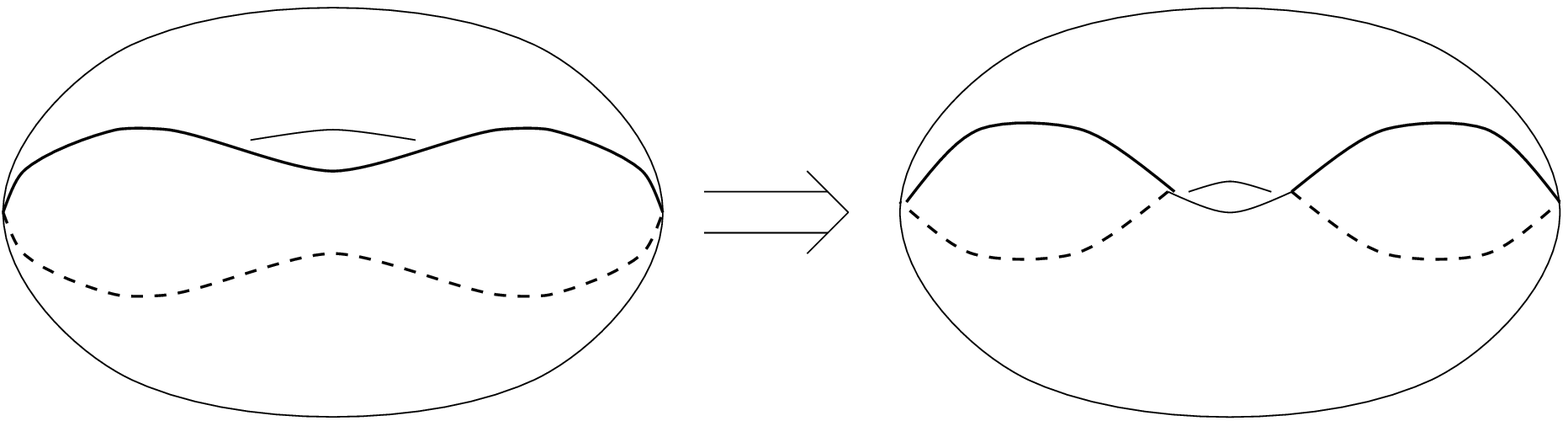}{\it Fig.4: A pinched surface and
   its transformation after $K_p$}{100mm}
\vskip 0.5cm

If $x,y\in S$ are two points that are distinct on the
world-sheet but $x=y$ at target space, we will 
use the notation $K_{xy}(S)$ for the surface that
is obtained by passing the infinitesimal string from $x$
to $y$.

Our suggestion for the surface equation is thus:
\bear
L_{\u\v}(x) &=& \px{\t}\lylx{}{\s_{\u\v\t}(x)} \nn\\
L_{\u\v}(x)\surf{S} &=& 
g^2 N \int_S dy_\u\wdg dy_\v \delta^{(6)}(y-x) 
\{\surf{K_{xy}(S)} -\inv{N}\surf{S}\} \label{SurfEqNA} 
\eear
The surface variables should also satisfy:
\be
\ep{\a\b\g\u\v\t} \px{\u}(x)\lylx{}{\s_{\a\b\g}(x)} \surf{S} = 0
\ee
Note that by naive scaling arguments the equations are
indeed scale invariant precisely in six dimensions, as was
expected.

In spite of the Bianchi identity, the surface functionals
cannot be of Stokes-type, i.e. expandable as:
\be
\sum_{r=0}^\infty \int_{S\times \cdots S}
 F_{\u_1\v_1,\dots,\u_r\v_r}(x^{(1)},\dots,x^{(r)})
\prod_{i=1}^r (dx^{(i)}_{\u_i}\wdg dx^{(i)}_{\v_i})
\ee
If this expansion were true, it would essentially mean that
$\surf{S}$ is given by a set of correlators
of points, i.e. by a {\em field theory}.

In fact, if $\surf{S}$ would be of Stokes-type, then (\ref{SurfEqNA})
could not be satisfied.
Unlike loops, where expanding 
$\px{\u}\lylx{}{\s_{\u\v}}$ in terms of the $F_{i_1\dots i_n}$ of the
Stokes-type expansion produces several terms which correspond 
to the non-abelian terms in QCD (see \cite{MigREV})
\be
\lylx{}{\s_{\u\v}(x)} : F_{\u_1\dots \u_n}(x_1,\dots,x_n)
\longrightarrow
\px{\lbr\u} F_{\v\rbr,\u_1\dots \u_n}(x,x_1,\dots,x_n)
+F_{\lbr \u\v\rbr,\u_1\dots \u_n}(x,x,\dots,x_n)
\ee
 the surface case would produce only  one term:\footnote{
I am grateful to A.A. Migdal for pointing this out,
as well as the following issue about perturbative solutions.}
\be
\lylx{}{\s_{\u\v\t}(x)} : F_{\u_1\v_1\dots \u_n\v_n}(x_1,\dots,x_n)
\longrightarrow
\px{\lbr\u} F_{\v\t\rbr,\u_1\dots \u_n}(x,x_1,\dots,x_n).
\ee
One can try to solve  (\ref{SurfEqNA}) perturbatively
in $g^2 N$. Let's put $N=\infty$ for simplicity.
Then, for the first iteration we would set $\surf{S} = 1$.
The next iteration would produce
\be
\surf{S} \sim \int_{S\times S}
 D_{\u\v,\t\s}(x-y)
(dx_\u\wdg dx_\v) (dy_\t\wdg dy_\s)
\label{Substit}
\ee
where $D_{\u\v,\t\s}(x-y)$ is the Abelian propagator.
Now, plugging this back in the RHS of (\ref{SurfEqNA})
we seem to be getting an abelian perturbative expansion.
The same procedure in QCD would also produce non-abelian
``gluon-interactions'' because plugging the substitution
in the LHS as well produces additional terms. But we saw
above that those terms are absent for surfaces. So we
seem to get only the abelian solution. This cannot be true
since the abelian solution satisfies the abelian
surface equation and not (\ref{SurfEqNA}). What goes
wrong is that when we plug (\ref{Substit}) in 
$\surf{K_{xy}(S)}$ we get a divergence. Here is precisely
where the ``stringy'' nature enters! In QCD we didn't get
a divergence because the contraction in the analogous
$\surf{K_{xy}(C)}$ involved only two {\em points} which
coincide. Here, the contraction in $\surf{S}$ indeed involves
two points in target space,
because the surfaces intersect transversely,
but on the world-sheet $K_{xy}(S)$ contains a whole loop that shrunk.
Indeed it seems like a ``string'' of zero length is propagating
from point $y$ on one sheet to point $x$ on
the other sheet.
The definition of the integral of the singular
propagator on $K_{xy}(S)$ cannot be regularized because it is
a limit of a shrunk loop and not just two points that coincide
on the world-sheet.

\subsection{Operator formalism}
Without any underlying Lagrangian, how can we tell whether
the set of equations (\ref{SurfEqNA}) corresponds to a consistent
quantum theory? We can try to find a set of operators
whose expectation values reproduce (\ref{SurfEqNA}).
We will construct these equations in a completely
``geometrical'' way. The operators will be labeled by surfaces
and locality will be implemented by commutation relations
that are non-zero only when two surfaces intersect.

In this way unitarity is easy to check but 6D Lorentz invariance
is not manifest.  As in QFTs, 6D Lorentz invariance becomes
apparent when Schwinger-Dyson equations for correlators are written.
In our case reproduction of (\ref{SurfEqNA}) will make Lorentz invariance
obvious.

In principle,
there might be a flaw in this argument because of world-sheet
gravitational anomalies. In constructing (\ref{SurfEqNA}) we
have defined the surfaces as {\em maps} from some world-sheet
to the 6D target-space and so we tacitly assumed that
there are no anomalies under reparameterization of the world-sheet.
In an operator formalism, such anomalies might cause
anomalies in the 6D Poincar\'e algebra. 
Nevertheless, we believe that as long as we are not including chiral
fermions or anti-self-dual fields, we will be safe.

We proceed to describe how the geometrical operator formalism
works in Yang-Mills and in the present case.

\subsubsection{Loop operators in QCD}
The loop representation of QCD was discussed in \cite{GamTri}.
Working in the temporal gauge
$A_0 = 0,$
the operators are  Wilson loops that lie on an equal-time
plane together with 
equal-time operators of the form
\bear
\lefteqn{
W_{i_1 i_2 \dots i_k} (C;\,\, x_1,\dots x_k) = 
}\nn\\
&& \tr {P\, e^{i \int_{C_{x_k x_1}} A \cdot dy}
  \dot{A}_{i_1}(x_1)
e^{i\int_{C_{x_1 x_2}} A \cdot dy}
  \dot{A}_{i_2}(x_2)
\cdots
e^{i\int_{C_{x_{k-1} x_k}} A \cdot dy}
  \dot{A}_{i_k}(x_k)
}
\nn
\eear
where $C_{xy}$ is the part of the loop $C$ from $x$ to $y$.

The commutation relations among the $W$ operators are
(see eqs (3.9 -- 3.10) of \cite{GamTri}):
\bear
\com{W(C)}{W(C')} &=& 0 \nn\\
\com{W(C)}{W_i(C';\,\,x)} &=& g\oint_C dy_i \dlu{3}(y-x)
\left\{ W(C_{0y}C'_{xx}C_{y0}) - \inv{N} W(C) W(C')\right\} \nn
\eear
and similar equations for an arbitrary number of insertions of
$\dot{A_i}$.

The equation of motion for $W$ that is
not self-intersecting is ($\hat{i_r}$ means
that $i_r$ is excluded):
\bear
{{d}\over {dt}}
W_{i_1 i_2 \dots i_k} (C;\,\, x_1,\dots x_k) &=&
\sum_{r=1}^k \int_{C_{x_{r-1} x_r}}
W_{i_1 \dots i_{r-1} j\, i_r \dots i_k}(C;\,\, x_1,
  \dots x_{r-1}, y, x_r \dots x_k)  dy_j
\nn\\ 
&+& g^2 \sum_{r=1}^k \px{j}(x_r)\lylx{}{\s_{i_r j}(x_r)}
W_{i_1 \dots \hat{i_r} \dots i_k}(C;\,\, x_1,\dots \hat{x_r} \dots x_k)
\nn
\eear
 For a loop $W$ that does self-intersect, say $x(s_1) = x(s_2)$,
we have to worry about the
normal ordering of the operators $A_i(x(s_1))$ and $\dot{A}_i(x(s_2))$.
Thus we get an extra term in the time derivative of $W$:
\be
{{d}\over {dt}}W = \cdots +
g\oint_{C\times C} \dlu{(3)}(y-x)
\left\{
W(C_{xy})W(C_{yx}) - \inv{N} W(C)
\right\}
dx_i dy_i
\label{LoopNORD}
\ee
Here $C_{xy}$ is the closed loop from $x$ to $y$ as in the loop 
equations.
Now we can check whether the equations of motion are consistent
by requiring
\be
{{d}\over {dt}}\com{{\cal O}_1}{{\cal O}_2}
=
\com{\dot{{\cal O}}_1}{{\cal O}_2}
+\com{{\cal O}_1}{\dot{{\cal O}}_2}
\ee
Unitarity requires (eq. (3.17-3.18) of \cite{GamTri}):
\bear
W(\bar{C}) = W^\dagger (C) 
\eear
and is supplemented by 
\be
|\bra\Psi |  W(C)  | \Psi\ket | \le 
\bra\Psi  | W(C_0) | \Psi\ket
\ee
where $C_0$ is the null loop and $\bar{C}$ is the loop $C$
with a reversed orientation.

The Hilbert space is formally the space of functionals of loops.
We can think of it as being regularized on a finite lattice.
Then, in order to correspond to $SU(N)$ Yang-Mills we have
to add the extra requirements when a link belongs to more than
$N$ loops (see \cite{MigREV} for details).
All the relations of this kind will make the Hilbert space finite.

\subsubsection{Surface operators in 6D}
Generalizing to $5+1$ dimensions we would have 
operators $\osurf{S}$ labeled by surfaces $S$ of 
various genera in 5D.
Those will commute among themselves, but we need to add
surfaces  with an arbitrary number of marked points.
Each point will be supplemented with a pair of indices.
This will be analogous to loops with field strength $\dot{A}_i$ 
insertions.
\be
\osurfi{i_1 j_1,\dots i_k j_k}{S;\,x_1,\dots, x_k}
\ee
The operator is antisymmetric with respect to interchange within
each pair (independently) $i_r j_r \rightarrow j_r i_r$.

The commutation relations among those operators will be nonzero
only in case one of the $x_i$-s of one operator lies on the 
surface of the other operator:
\bear
\lefteqn{
\com{\osurfi{i_1 j_1,\dots, i_k j_k}{S;\,x_1,\dots,x_k}}{
\osurfi{i_1' j_1',\dots, i_l' j_l'}{S';\,x_1',\dots,x_l'}}
}\nn\\ && 
 = g\sum_{r=1}^k \int_{S'} 
\left(
\osurfi{i_1 j_1,\dots \widehat{i_r j_r} \dots, i_k j_k,
i_1' j_1',\dots, i_l' j_l'}{K_{y,x_r}(S,S');\,x_1,\dots,\widehat{x_r},
\dots, x_k,x_1',\dots,x_l'} \right.\nn\\
&-&
\left.
\inv{N}
\osurfi{i_1 j_1,\dots \widehat{i_r j_r} \dots, i_k j_k}{S;\,
x_1,\dots,\widehat{x_r},\dots, x_k}
\osurfi{i_1' j_1',\dots, i_l' j_l'}{S';\, x_1',\dots,x_l'}
\right)
\dlu{(5)}(x_r - y) dy_{i_r}\wdg dy_{j_r} 
\nn\\
&-&
g\sum_{r=1}^l \int_{S} 
\left(
\osurfi{i_1 j_1,\dots, i_k j_k,
i_1' j_1',\dots \widehat{i_r' j_r'} \dots, i_l' j_l'}{K_{y x_r'}(S,S');
\,x_1,\dots, x_k,x_1',\dots,\widehat{x_r'},\dots,x_l'} 
\right.\nn\\
&-&
\left.
\inv{N}
\osurfi{i_1 j_1,\dots, i_k j_k}{S;\, x_1,\dots,x_k}
\osurfi{i_1' j_1',\dots \widehat{i_r' j_r'} \dots, i_l' j_l'}{
S';\, x_1',\dots,\widehat{x_r'},\dots,x_l'}
\right)
\dlu{(5)}(x_r - y) dy_{i_r'}\wdg dy_{j_r'} 
\nn\\ &&\label{SurfCom}
\eear
where $\widehat{x_r}$ means that $x_r$ is excluded.
$K_{y y'}(S,S')$ is the operation of gluing two surfaces
$S$ and $S'$ at an intersection point $S\ni y = y'\in S'$
which was described above (in Figs 2-4).

The time evolution for a single insertion is given by:
\bear
\lefteqn{
{d\over {dt}} \osurfi{ij}{S;\,x} =
}\nn\\ &&
\int_S \osurfi{ij,i'j'}{S;\,x,y} dy_{i'}\wdg dy_{j'}
+\px{k}(x)\lylx{}{\s_{ijk}(x)}\osurf{S} \nn\\
&+& g^2\int_{S\times S} \dlu{(5)}(y-y')
\left(\osurfi{ij}{K_{y,y'}(S);\,x} -\inv{N}\osurfi{ij}{S;\,x}\right)
(dy_{k}\wdg dy_{l}) (dy'_{k}\wdg dy'_{l})
\label{SurfTiEv}
\eear
The last line is similar to the last line in (\ref{LoopNORD}).
The case of many insertions is similar.

We have to check that these rules are consistent with
the chain rule for differentiation:
\be
{{d}\over {dt}}\com{{\cal O}_1}{{\cal O}_2}
=
\com{\dot{{\cal O}}_1}{{\cal O}_2}
+\com{{\cal O}_1}{\dot{{\cal O}}_2}
\ee
The only non-trivial case is when the surface corresponding
to ${\cal O}_1$ intersects itself at $x=y$ and the surface
corresponding to ${\cal O}_2$ intersects the self-intersecting
surface at the same point $z=x=y$.
In this case it is a subtle question of regularization.
We will discuss a similar case in section (4.3).

The unitarity requirement will be that orientation reversal
of the world-sheet is the same as complex conjugation.

\section{The anti-self-dual tensionless string}
The tensionless strings that are derived from the 
(10D/11D) string theory constructions couple to an
anti-self-dual 2-form $B_{\u\v}^{(-)}$.
Supersymmetry in 6D allows only anti-self-dual 2-forms,
if we do not wish to include the graviton.

The previous section dealt with a theory that has
a full 2-form (and no supersymmetry, of course).
In what follows we will discuss the anti-self-dual case.

\subsection{The abelian anti-self-dual two-form}
The abelian anti-self-dual surface equations are
obtained from the free anti-self-dual two-form theory
in 6D.
We consider only surfaces that are all in a constant time
hyper-plane.
The operators of the theory satisfy the relations
\bear
H_{ijk}^{(-)} &=& \px{\lbr k}B_{ij\rbr}^{(-)} \nn\\
\px{0}B_{lm}^{(-)} &=& \inv{3!}\ep{ijklm}H_{ijk}^{(-)} 
\nn\\
\com{H_{ijk}^{(-)}(x)}{B_{lm}^{(-)}(y)} 
 &=& i\ep{ijklm} \dlu{(5)}(x-y) \nn
\eear
Second quantization is done by writing out the
second quantized operator for the full (self-dual plus
anti-self-dual) theory with Lagrangian $\int |dB|^2 d^6x$
and then discarding the self-dual part of the resulting
$B_{\u\v}$ operator.

The surface operators are
\be
\osurf{S} =\,\, :exp\{i g\int_S B^{(-)}\}:\,\,
 =\,\, :\exp\{i g\int_{M:S = \partial M} H\}:
\ee
Here $M$ is any manifold whose boundary is $S$.
If $S$ and $S'$ do not intersect then they commute.
They satisfy (see also \cite{Leal}):
\be
\osurf{S}\osurf{S'} = e^{ig^2\lnk{S,S'}}
\osurf{S'}\osurf{S}
\ee
where $\lnk{S,S'}$ is the integer linking number of $S$ and $S'$,
In principle, $g$ could have any value, but if $\osurf{S}$ 
is to describe the action of some stringy physical objects,
they must satisfy Dirac's quantization condition
\be
g^2 = 2\pi
\ee
and $\osurf{S}$ commutes with $\osurf{S'}$ if $S$ and $S'$
do not intersect.

Since 
\be
\lylx{}{\s_{ijk}(x)}\lnk{S,S'} = 
\ep{ijklm}\int_{S'}\dlu{(5)}(x-y) dy_l\wdg dy_m
\ee
we find for $g^2 = 2\pi$:
\be
\com{\lylx{\osurf{S}}{\s_{ijk}(x)}}{\osurf{S'}}
=
2\pi i\ep{ijklm}\int_{S'}\osurf{S,S'}\dlu{(5)}(x-y) dy_l\wdg dy_m
\ee
where $\osurf{S,S'}$ is the normal-ordered product.

 From the equations of motion we find the time derivative
for surfaces that do not intersect themselves:
\be
{d\over {dt}}\osurf{S} = 
\ep{ijklm}\int_S \lylx{\osurf{S}}{\s_{ijk}(x)} dx_l \wdg dx_m 
\ee

\subsection{The non-abelian equations}

What could be a possible non-abelian generalization of
the operator algebra of the abelian case?

We should write down equations for
\be
{d\over {dt}}\osurf{S} = \cdots
\ee
as well as commutator relations.

Because of anti-self-duality we expect something like
\be
{d\over {dt}}\osurf{S} =
\int_S \ep{ijklm} \lylx{\osurf{S}}{\s_{ijk}} dx^l\wdg dx^m
+{\mbox{(contact terms)}}
\ee
The ``contact-terms'' should appear only for a self-intersecting
loop.
In fact, we should be able to define a Hamiltonian.
Although the operators are in general non-local it might be
the case that there exist a local energy momentum tensor.\footnote{
This was suggested by E. Witten.}
In this case we can write the Hamiltonian as
\be
\Ham = \int \Ham(x) d^5x
\ee
In the abelian case 
\be
\Ham(x) = \hlf (H_{ijk}^{(-)}(x))^2
\ee
where $H_{ijk}^{(-)}$ is the (anti-self-dual) field strength.
$\Ham(x)$ can be defined as a limit of  a second derivative
of a small spherical surface containing $x$:
\be
\hlf\lim_{S\rightarrow \{x\}} 
\llylxx{}{\s_{ijk}(x)}{\s_{ijk}(y)}\osurf{S}
\ee
here as $S$ shrinks to the point $x$, $y\rightarrow x$ too.
Let us assume that the same expression holds in the non-abelian
case as well. After all, in Yang-Mills theory a similar
expression (with the addition of a $\tr{E_i^2}$ describes
the Hamiltonian too).
The problem is to find a consistent set of equations for products
of operators.

As in the field theory case, where it is more convenient
to write the commutation relation 
$\com{H_{ijk}^{(-)}(x)}{B_{mn}^{(-)}(y)}$
than $\com{B_{jk}^{(-)}(x)}{B_{mn}^{(-)}(y)}$ we will
make a guess for the commutation relation:
\be
\com{\lylx{\osurf{S}}{\s_{ijk}(x)}}{\osurf{S'}}
= {\mbox{(contact terms)}}
\ee
we will then have to check the identities
\bear
\lylx{}{\s_{lmn}(y)}\com{\lylx{\osurf{S}}{\s_{ijk}(x)}}{\osurf{S'}}
&=&
\lylx{}{\s_{ijk}(x)}\com{\lylx{\osurf{S}}{\s_{lmn}(y)}}{\osurf{S'}},
\qquad x,y\in S
\label{Id1}\\
\lylx{}{\s_{lmn}(y)}\com{\lylx{\osurf{S}}{\s_{ijk}(x)}}{\osurf{S'}}
&=&
\lylx{}{\s_{ijk}(x)}\com{\osurf{S}}{\lylx{\osurf{S'}}{\s_{lmn}(y)}},
\qquad x\in S,\qquad y\in S'
\label{Id2}
\eear
as well as the Jacobi identity
\bear
\com{ \lylx{\osurf{S}}{\s_{ijk}(x)} }{ 
\com{ \lylx{\osurf{S'}}{\s_{lmn}(y)}}{\osurf{S''}} }
&-&
\com{ \lylx{\osurf{S'}}{\s_{lmn}(y)} }{ 
\com{ \lylx{\osurf{S}}{\s_{ijk}(x)} }{\osurf{S''}} }
\nn\\
&& =
\com{
\com{\lylx{\osurf{S}}{\s_{ijk}(x)} }{ 
     \lylx{\osurf{S'}}{\s_{lmn}(y)} }}{\osurf{S''}}
\label{Jacob1}
\eear

Our guess for the commutator is
\bear
\com{\lylx{\osurf{S}}{\s_{ijk}(x)}}{\osurf{S'}}
&=&
\ep{ijklm}
\int_{S'} (\osurf{K_{xy}(S,S')} -\inv{N}\osurf{S,S'})
\dlu{(5)}(x-y) dy_l\wdg dy_m
\label{SurfComASD}
\eear
Here we assume that $S$ and $S'$ {\em do not intersect at points
other than $x$}. If they do there are more contact terms
that have to be taken care of by including more 
$\lylx{}{\s_{ijk}}$ derivatives.

They seem to satisfy (\ref{Id1},\ref{Id2},\ref{Jacob1}).
Of course, the only non-trivial check is the singular contribution
when $x=y$ in those equations.
The check for (\ref{Id1}-\ref{Id2}) only involves derivatives
of the $\delta$-functions and not the special nature of $K_{xy}$.
Since (\ref{Id1}-\ref{Id2}) are certainly satisfied for the abelian
free field case, they continue to do so for the non-abelian case 
as well.
On the other hand (\ref{Jacob1}) depends on the specific nature
of the new term with $K_{xy}$. Thus, the precise regularization
here is important. We can define $K_{xy}(S,S')$
with a world-sheet that looks locally like a cylinder at $x=y$,
parameterized by $-\infty<r<\infty$ and $0\le \theta\le 2\pi$
such that the circle at $r=0$ maps to $x=y$ whereas the part $r<0$
maps to $S$ and the part $r>0$ maps to $S'$. We can then
use this world sheet for the RHS in 
$\com{\lylx{\osurf{S'}}{\s_{lmn}(y)}}{\osurf{S''}}$ in (\ref{Jacob1}).
Then we have to commute this with $\lylx{\osurf{S}}{\s_{ijk}(x)}$.
Using (\ref{SurfComASD}) again, we would get some terms with a
delta function $\dlu{(5)}(x-\xi)$ where $\xi$
is a coordinate on $K_{y}(S',S'')$. If we regularize it by
a Gaussian $e^{-|x-\xi|^2}$ for example, then we pick
contributions from both $S'$ and $S''$. Thus, in (\ref{Jacob1})
the first term picks contributions from $S,S'$ and from $S,S''$,
the second term from $S',S''$ and $S,S''$ and the RHS from 
$S,S''$ and $S',S''$. So they indeed seem to cancel.
I do not know if there are any subtleties in the regularization
that invalidate this argument.

\section{Supersymmetry}
The $\SUSY{1}$ supersymmetry multiplets 
in 6D are:\footnote{
This table is taken from \cite{DMW}.}
\be
\begin{array}{ll}
Supergravity~Multiplet
~~~~~&G_{MN},\Psi^{A+}_{M},B^+{}_{MN}\\
Tensor~Multiplet~~~~~~~~~~~~&B^-{}_{MN},\chi^{A-},\Phi\\
Hypermultiplet~~~~~~~&\psi^{a-},\phi^{\alpha}\\
Yang-Mills~Multiplet~~~~~~~~&A_{M},\lambda^{A+}
\end{array}
\ee
All spinors are symplectic Majorana--Weyl.
The two-forms $B^+{}_{MN}$ and $B^-{}_{MN}$
have three-form field strengths that are self-dual and
anti-self-dual, respectively.

 For a super-symmetric theory of surface operators, 
we need to add operators labeled by surfaces with marked
points. These would be generalizations of loop
operators in 4D QCD like
\be
\tr{P \lam^\a(x) e^{i\oint_{C_{xx}} A_k dx_k}}
\ee
where $\lam$ is the gluino in Super-Yang-Mills
(see also \cite{ItoTak} for a related discussion).
In 4D $\SUSY{2}$ QCD there are also bosonic fields
in the adjoint representation and so we need
to add loops with marked points that correspond
to the bosonic fields as well.

In the surface case, we will try to do the same.
We add surfaces with marked points. Each point
corresponds either to a bosonic $\Phi$ field
or to a fermionic $\chi^{(-)}$ field of the
tensor multiplet. 
Each fermionic marked point is supplemented with
a spinor index.
Each bosonic marked point can be either a $\Phi$
or a $\dot{\Phi}$ insertion.
In the 4D QCD case, the commutation relations
among the different operators are all derived
from the commutation relations among the fields.
 For example, when a $\Phi(x)^a\tau^a$ insertion
on one loop meets a $\dot{\Phi}^b(y)\tau^b$
insertion on another loop (when $x=y$)
we get a term
\be
\com{(\Phi(x)^a\tau^a)^r_s}{(\dot{\Phi}^b(y)\tau^b)^p_q}
\sim
\delta(x-y) (\delta^r_q\delta^s_p - \inv{N}\delta^r_s\delta^p_q)
\ee
here $p,q,r,s$ are $SU(N)$ indices and $\tau^a$, $a=1\dots N^2-1$
are generators of $SU(N)$.
In the QCD case, these rules imply the same kind of
geometrical operation on the loops, so we guess that in 6D
whenever a $\Phi$ insertion meets a $\dot{\Phi}$ insertion
on another surface, the commutator will include a term
like the RHS of (\ref{SurfComASD}).
Surface operators with an odd number of fermionic
insertions will be anti-commuting and their anti-commutator
will contain the appropriate spinor index matrix.

\section{Adiabatic arguments for large $N$}
We would like to learn more about the large $N$ limit
of tensionless string theories, directly from the 10D/11D string
theory construction.

We will discuss two cases. 
The first case will be 
the $\SUSY{2}$ theory, obtained as type-IIB on an $A_{N-1}$
singularity which is modeled by $\MC{2}/\IZ_N$ where the
$\IZ_N$ acts as
\be
\IZ_N: (z_1,z_2)\longrightarrow (e^{{{2\pi i}\over N}} z_1,
e^{-{{2\pi i}\over N}} z_2)
\ee
(it has to be mentioned that we need the conifold
and not the orbifold,
which defer by a $\theta$-angle \cite{AspENH}).

The second case is the small $E_8$ instanton at instanton
number $N$. Here we will use the effective field-theory
solution of Callan-Harvey-Strominger (the symmetric
five-brane) in the large $N$ limit.

We will find that in both cases
an extra dimension decompactifies.

\subsection{Type-IIB on $A_{N-1}$ at large $N$}
Parameterizing the conifold as:
\be
\IZ_N: (z_1,z_2)\longrightarrow (e^{{{2\pi i}\over N}} z_1,
e^{-{{2\pi i}\over N}} z_2)
\ee
we can think of the space as an angle of ${{2\pi}\over N}$
in the $z_1$-plane, over each point of which we have a $z_2$
plane. At the two boundaries of the ${{2\pi}\over N}$ angle
the $z_2$ planes are identified with a twist of $-{{2\pi}\over N}$.
When $N$ is large, the two sides of the angle in the $z_1$
plane are almost parallel. Moreover, if we restrict to the
vicinity of the origin in the $z_2$ plane, the $-{{2\pi}\over N}$
twist is also small. Thus, we can think of the close vicinity
of the $z_2$ origin as part of a compactification on $\MS{1}$
of radius ${R\over N}$, where $R$ is the distance to the origin
in the $z_1$ plane (see Fig 5).
\vskip 0.5cm
\putfig{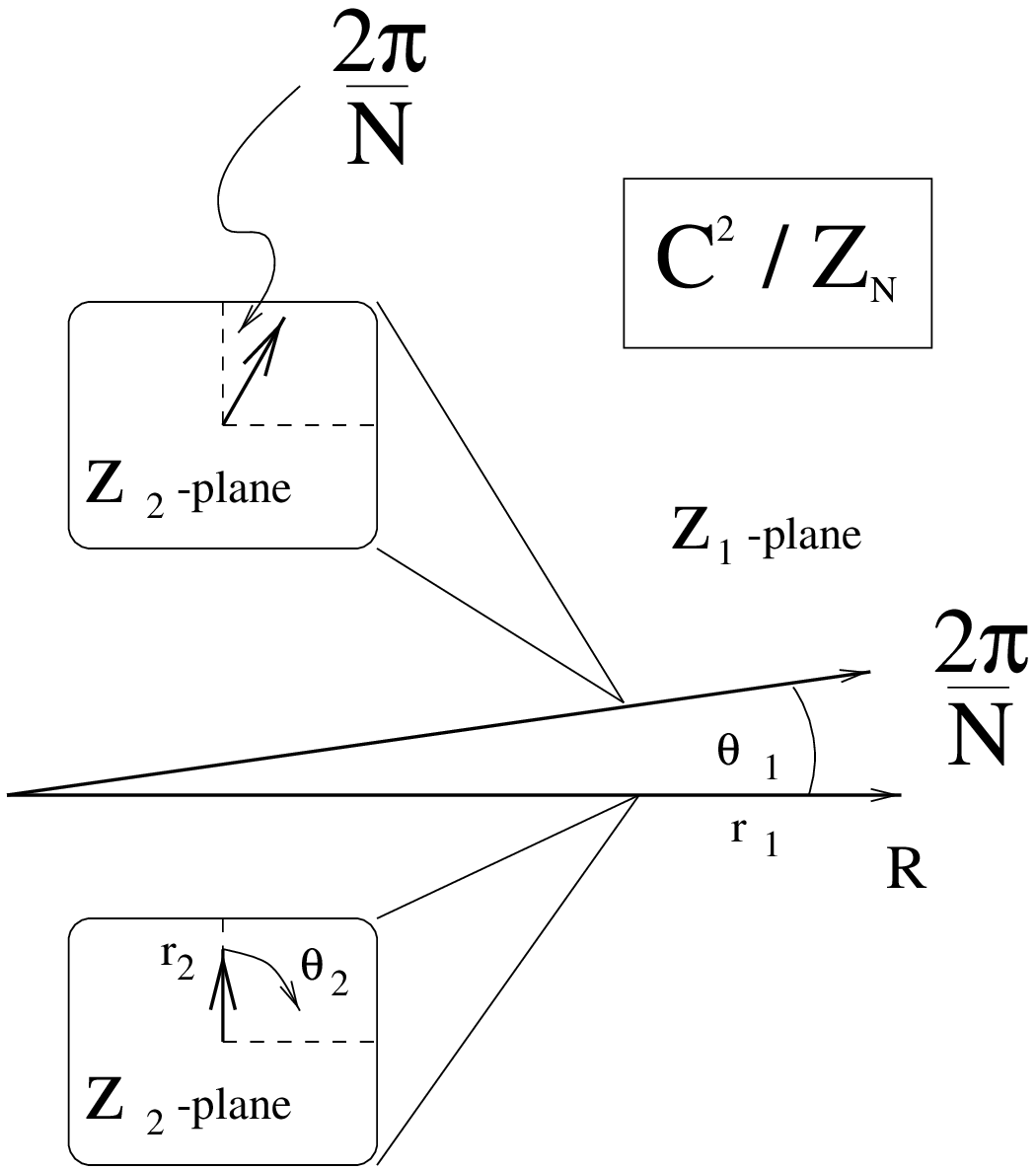}{\it Fig.5: Locally, the orbifold $\MC{2}/\IZ_N$
   looks like a circle of radius ${R\over N}$ times a twisted $z_2$
   plane.}{100mm}
\vskip 0.5cm
In the adiabatic approximation, we will T-dualize type-IIB 
on this construction to, approximately, type-IIA on a circle
of radius ${N\over R}$ at the vicinity of the origin.
The circle grows to infinity as $R\rightarrow 0$.
T-duality also
multiplies the 10D coupling constant by ${{N}\over {R}}$
so that we have a strongly coupled type-IIA which grows
an 11th dimension \cite{WitVAR}.

Let us check precisely the transformations and the
region of applicability of the adiabatic argument, i.e.
the region where the curvature is small.

We parameterize the conifold by
$r_1,\theta_1, r_2,\theta_2$ with
\be
0\le r_1,\qquad
0\le r_2,\qquad
0\le\theta_2\le 2\pi,\qquad
0\le\theta_1\le {{2\pi}\over N}
\ee
$r_1,\theta_1$ are polar coordinates for the $z_1$-plane
and $r_2,\theta_2$ are polar coordinates for the $z_2$-plane.
The $\theta_1$ and $\theta_2$ coordinates live
on a torus. The original type-IIB metric is
\be
ds^2 = dr_1^2 + dr_2^2 + r_1^2 d\theta_1^2 
+r_2^2 d\theta_2^2
\ee
Now we define
\be
\theta_1 = {y\over N}, \qquad
\theta_2 = x+ {y\over N}
\ee
so $x,y$ live on a square torus of unit size, and the metric is
\be
ds^2 = dr_1^2 + dr_2^2 + {{r_1^2+r_2^2}\over {N^2}} dy^2 
+r_2^2 dx^2 + {{2r_2^2}\over N}dx dy
\ee
T-duality on a torus with metric $G_{11}, G_{12}, G_{22}$
replaces the metric with 
\bear
G_{11}' &=& G_{11} \nn\\
G_{12}' &=& 0 \nn\\
G_{22}' &=& {{G_{11}} \over {G_{11} G_{22} - G_{12} G_{12}}} \nn
\eear
and now there is also a $B_{12}$ field
\be
B_{12} = {{G_{11} G_{12}} \over {G_{11} G_{22} - G_{12} G_{12}}}
\ee
So after T-duality in the $y$ direction
we get a metric (using the adiabatic approximation):
\be
dr_1^2 + dr_2^2 + r_2^2 dx^2 + {{N^2}\over {r_1^2}} dy^2
\ee
and a $B$ field
\be
B_{xy} = {{N r_2^2}\over {r_1^2}}
\ee
The coupling constant becomes:
\be
\lam_A = \lam_B {{N}\over {r_1}}
\ee
Transforming to M-theory we find the metric
\bear
ds^2 
&=&
 {9\over {16}}\lam_B^{-2/3}N^{-2/3} d(r_1^{4/3})^2
+\lam_B^{-2/3} N^{-2/3} r_1^{2/3} dr_2^2
+\lam_B^{-2/3} N^{-2/3} r_1^{2/3} r_2^2 dx^2 \nn\\
&+&\lam_B^{-2/3} N^{4/3} r_1^{-4/3} dy^2
+\lam_B^{4/3} N^{4/3} r_1^{-4/3} (dx^{11})^2 \nn
\eear
and
\be
A_{xy11} = {{N r_2^2}\over {r_1^2}}
\ee
Defining
\be
u = \lam_B^{-1/3}N^{-1/3}r_1^{4/3}, \qquad
v = \lam_B^{-1/4} N^{-1/4} r_2
\ee
we obtain
\bear
ds^2 &=& {9\over {16}} du^2 
+u^{1/2} dv^2
+u^{1/2} v^2 dx^2
+\lam_B^{-1} N u^{-1} dy^2
+\lam_B N u^{-1}(dx^{11})^2 \nn\\
A_{xy11} &=&  v^2 u^{-3/2} \nn
\eear
An estimate of the tensor $R_{mnpq}R^{mnpq}$
gives, for fixed $\lam_B$, the conditions (of small curvature):
\be
u^2 \gg 1,\qquad u^{1/2}v^2 \gg 1
\ee
Similarly, a calculation of $|dA|^2$ gives the condition
(of small gradient):
\be
\inv{N^2 u^2}\ll 1,\qquad 
{{v^2}\over {N^2 u^{5/2}}} \ll 1
\ee
which together give
\be
1\ll u,\qquad
u^{-1/2} \ll v^2 \ll N^2 u^{5/2} \label{CondCurv}
\ee
The conditions for decompactification in the large $N$ limit
 were that the
sizes of the $dy$ and $dx^{11}$ directions be large.
This gives
\be
u\ll N
\ee
which is consistent with (\ref{CondCurv}) for a certain
range of $u$.

One might be worried about the fact that this range corresponds
to large $r_1$ in the original variables.
This range stops approximately where $r_1\sim N^{1/4}$.
However, the phenomenon that we are looking for is generic,
in the sense that it should correspond to ``something new''
even if we take type-IIB on an infinite $\MR{4}$
divided by $\IZ_N$. The extra dimensions that decompactify
close to the singularity have sizes much larger than
the distance to the origin.

To be more precise, if we start with type-IIB on $\MR{4}/\IZ_{N}$
for $N$ finite,
a low-energy observer in 10D would describe the situation as
having type-IIB super-gravity in the 10D bulk which interacts
with a tensionless string theory on the 6D singularity.
From the above formulae, we learn that for $u\sim N^\epsilon$
(for a small positive $\epsilon$) the extra uncompactified
dimensions have sizes of order $N^{\hlf-\hlf\epsilon}$ in M-theory
units. This will correspond to Kaluza-Klein masses of order
$N^{-1/2+\epsilon/2}$ in these units.
Thus, if we probe energies which are much smaller than
$N^{-\epsilon}$ but not smaller than $N^{-1/2+\epsilon/2}$,
we will think that the phenomenon that we are observing is
coming directly from the singularity ($u$ smaller than
our sensors) but we will observe the extra dimensions.

Another thing that we notice about the metric is that
even if we take the limit of weak $\lam_B$ or strong $\lam_B$,
at least one of the dimensions $dy$ or $dx^{11}$ will be large.

On the other hand,
it could also be that the states of masses $N^{-1/2}$ have
nothing to do with the tensionless string theory but are
only related to type-IIB on $\MR{4}/\IZ_N$.
However, we note that the applicability of the
 argument ceased because curvatures became large and not
because the size of the extra dimensions became small.
The fact that the other tensionless string theory
related to small $E_8$ instantons also exhibits
a similar phenomenon, as we will discuss shortly,
 might be considered as supporting
evidence for relating the phenomenon to tensionless
string theory.

\subsection{A check on type-IIA on $A_{N-1}$}
Let us see what would happen to a similar
argument for type-IIA on an $A_{N-1}$ singularity.
The argument should not work in this case, of course,
since type-IIA on $A_{N-1}$ produces an enhanced gauge
symmetry  and not an exotic tensionless string theory.

In this case we would get again
\be
ds^2 = dr_1^2 + dr_2^2 + r_2^2 dx^2 + {{N^2}\over {r_1^2}} dy^2
\ee
and a $B$ field
\be
B_{xy}^{(NS)} = {{N r_2^2}\over {r_1^2}}
\ee
The coupling constant becomes:
\be
\lam_B = \lam_A {{N}\over {r_1}}
\ee
But now we are in type-IIB so we have to use the S-duality
to transform to a weakly coupled theory.
\be
\lam_{\mbox{new}} = \inv{\lam_{\mbox{old}}}
\ee
This is accompanied by a rescaling
\be
g_{\mbox{new}} = \inv{\lam_{\mbox{old}}} g_{\mbox{old}}
\ee
so

\be
ds^2 = \lam_A^{-1}N^{-1} r_1 dr_1^2
+\lam_A^{-1}N^{-1} r_1 dr_2^2 
+\lam_A^{-1}N^{-1} r_1 r_2^2 dx^2
+{N \over {r_1}} dy^2
\ee
together with a coupling constant of 
\be
\lam = \lam_A^{-1}N^{-1} r_1
\ee
and an RR field
\be
B_{xy}^{(RR)} = {{N r_2^2}\over {r_1^2}}.
\ee
Now we have to T-dualize again.
The resulting metric is:
\be
ds^2 = 
\lam_A^{-1}N^{-1} r_1 dr_1^2
+\lam_A^{-1}N^{-1} r_1 dr_2^2
+\lam_A N r_1^{-1} r_2^{-2} dx^2 + {N \over {r_1}} dy^2
\ee
An estimate of the tensor $R_{mnpq}R^{mnpq}$
gives, for fixed $\lam_B$, the conditions (of small curvature):
\be
{{\lam_A N} \over {r_1^3}} \ll 1,
{{\lam_A N} \over {r_1 r_2^2}} \ll 1
\ee
So, for the applicability of the argument, the size of the $x$
direction has to be small. On top of that, the size of the $y$
direction is always smaller that $\sim N^{1/3}$ and so
is less than $r_1$. Thus, there is no decompactification
in this case as it should.

\subsection{The symmetric five-brane at large $N$}
The symmetric five-brane solution at instanton number $N$
is given by \cite{CHS}:
\bear
ds^2 &=& e^{2\phi}dx^2 + dy^2 \nn\\
e^{2\phi} &=& e^{2\phi_0} + {Q\over {x^2}} \nn\\
H &=& -Q\epsilon \nn
\eear
where $ds^2$ is the 10D metric, $x$ are the 4D coordinates (where
the instanton lives), $y$ are the 6D coordinates where the tensionless
string theory is observed,
$\phi$ is the dilaton field, $\phi_0$ is its value at 4D infinity,
$Q=N\a'$ is the charge which is proportional to the instanton
number $N$, $H$ is the 3-form field strength and $\epsilon$
is a 3-form constant in the directions tangent to the radius in 4D.

The 4D length scale is governed by $\sqrt{N\a'}$
so for large $N$ the fields vary slowly.
In fact, the curvature is proportional to $\inv{N\a'}$ and is small
compared to the string scale.
If we are dealing with $E_8$ instantons, we can think of locally
replacing the heterotic $E_8$ theory with dilaton $\phi$ by
M-theory on $\MS{1}/\IZ_2$ with size \cite{WitVAR,HorWit}:
\be
r = e^{2\phi/3} \sim {{N^{1/3}}\over {x^{2/3}}}
\ee
and the 4D metric is rescaled to 
\be
ds^2 \rightarrow e^{-2\phi/3}ds^2 = e^{4\phi/3}dx^2 
\sim N^{2/3}{{dx^2}\over {x^{4/3}}}
\ee
We see that a 7th dimension becomes uncompactified as before.

The M-theory metric is given by:
\be
(\lam_0^2 +{N\over {x^2}})^{4/3} dx^2 +
(\lam_0^2 +{N\over {x^2}})^{2/3} (dx_{11})^2
\ee
Neglecting $\lam_0$, an estimate for the M-theory curvature yields:
\be
|x| \ll N^2
\ee
which is certainly satisfied at the region of decompactification
\be
1\ll {N\over {x^2}} 
\ee

\subsection{Remarks on the surface operators and a continuous spectrum}
What are the implications of the decompactification?
We saw that a region close to the singularity ``grows''
an extra dimension. We cannot of course extrapolate
to the $A_{N-1}$ singularity itself because the curvature
there becomes large. It might be that something additional
``lives'' at the singularity but whatever it is, it seems
that it will have to interact with the extra decompactified
dimension.

The six-dimensional tensionless string theory
is scale-invariant. Thus, if it has any stable massive 1-particle states,
the spectrum has to be continuous.
The previous calculations suggest that as $N\rightarrow\infty$
there is a region close to the singularity where the curvature
is small but there is an extra dimension of size at least $O(N^{1/2})$
in M-theory units.

I would guess that at least Kaluza-Klein states of the gravitons exist
and are much lighter than the Planck scale.
Since the spectrum cannot be discrete it will follow that
there is a continuum of massive particles (The states that
are lighter than $O(\inv{N^{1/2}})$ will have 
to come from the region much closer to the singularity 
where the above approximations are not valid).

On the other hand, as was mentioned above, it might
be that those states of mass $O(\inv{N^{1/2}})$ have nothing
to do with tensionless string theory but exist in
type-IIB on $\MR{4}/\IZ_N$ as massive states that become
massless only in the large $N$ limit.

Even if the states are related to the large $N$ limit of
the tensionless string theory itself,
I do not know if these arguments hold for finite $N$ as well,
since it might be that the states discussed above are unstable for
finite $N$.

Can a surface operator formalism incorporate a continuous spectrum?

Let us consider the correlator  of two identical closed surfaces
of size $R$ located at points $x$ and $y$ in 6D.
If we are allowed to assume a similar behaviour as that
of large $N$ QCD, their correlator would behave as
\be
\ev{\osurf{S_x(R)}\osurf{S_y(R)}} \sim O(\inv{N^2}).
\ee
The tensionless string theory is assumed to be scale invariant
in addition. We therefore expect
\be
\ev{\osurf{S_x(R)}\osurf{S_y(R)}} \sim \inv{N^2} f({{|x-y|}\over R})
\ee
A continuous spectrum could be obtained if the function $f$
contains a piece like $f(x)\sim \cdots +e^{-cx}$.
Then, an operator $\osurf{S_x(R)}$ would create a state
of mass proportional to its size $R$.

\section{Discussion}

This paper dealt with two different aspects of
tensionless string theories in the large $N$ limit.
The first has to do with the variables of the theory
and the second concerns space-time.

In the first approach we suggested that surface-equations
in 6D analogous to loop-equations in 4D might describe
a consistent quantum theory which could be related to
tensionless string theories.
Here the motivation is to study the tensionless string theories
in 6D independently from the framework of 10D/11D string
theory.
Since the (10D/11D) string theory constructions are
non-perturbative one has to rely heavily on BPS arguments.
An independent construction of 6D tensionless strings
might encompass a larger class of theories, maybe even
without super-symmetry, and might shed more light 
on the non-perturbative superstring-theory phenomena.

The other approach was to try to deduce certain features
of the 6D theories using what we know about the 
non-perturbative superstring theory constructions.
The large $N$ limit allows for an adiabatic application
of duality transformations where we end up with more
than six ``large'' dimensions. The extra dimensions have
sizes proportional to $N^{1/2}$.
This has implications on the observed 6D spectrum of states.
There is however a caveat -- we couldn't entirely rule out the
possibility that those states might
just come from the large $N$ limit of type-IIB on $\MR{4}/\IZ_N$
and not from the tensionless string theory.

Returning to the question of what are the variables of
tensionless strings,
we suggested that within tensionless string theories it might
be possible to define correlators of operators labeled by surfaces
and we tried to guess what kind of local relations those
correlators might satisfy.

In 4D QCD the Wilson loops represent trajectories of charged
matter. What do the surfaces in the 6D theory represent?

We have concentrated on the 6D analog of pure Yang-Mills
theory so the question arises what would be the 6D analog
of the addition of matter.

It was argued in \cite{GanHan,SWCSD,WitPMF} that a heterotic
string vacuum in 6D with a zero size $E_8$ instanton in the
compactified dimensions can lead to tensionless strings
with an $E_8$ current algebra on their ``world-sheets''.
(There is no known world-sheet formulation of these tensionless strings,
so what we really  mean is
that once we turn on VEVs that
increase the tension we can reach  a theory with a stringy solitonic
object and an $E_8$ current algebra living on it.)
The candidates for ``stringy-matter'' thus seem to be
such strings with CFTs on their world-sheets.
Indeed, in the above example, upon compactification to 4D on a torus
with appropriate $E_8$ Wilson loops the new types of strings
become hyper-multiplets in the fundamental representation of the
gauge group.

In QCD, quarks are introduced by adding open Wilson loops
which correspond to a quark on one end and an anti-quark at the other end.
Going back to 6D, we will include open
surfaces as well, with boundaries  that have the quantum numbers
of the ``matter strings'' (e.g. a state in the chiral $E_8$
current algebra for the example discussed above).

However, we expect that not all possible CFTs should be allowed.
In \cite{GanHan} cancelation of world-sheet anomalies
required a CFT with $c-\bar{c} = 8$.
In the surface-equations, we have used a world-sheet
to define the surface (this was not the world-sheet of the
tensionless string, just of the operator). We have
assumed invariance under reparameterization of this world-sheet.
As long as we have no chirality in the theory -- P invariance
will probably prevent such anomalies.
However such anomalies should arise for the theories with
the anti-self-dual string. I do not know if this can be seen
in the surface-equations -- but this is a necessary check
on the equations. Maybe the anomalies arise as failure
of 6D Lorentz invariance, in the operator formalism.

Though the critical dimension for the surface equations
is six, they can be defined in lower dimensions as well.

Independently from the surface equations, analogy with
2D QCD might suggest that at a low enough dimension,
 3D maybe, there will be  no propagating degrees of freedom.
Maybe at those dimensions, the large $N$ limit can be
described by a topological membrane theory,
like 2D QCD can be described by a topological string theory
\cite{QCDTD}.

Finally, if indeed tensionless string theories in the large
$N$ limit grow a large 11th dimension, as we calculated in section (6),
then a better understanding of tensionless string theories
might be relevant for M-theory.

\section*{Acknowledgments}
I wish to thank K. Dienes, 
M. Flohr, A. Hanany, I. Klebanov, A.A. Migdal, S. Ramgoolam,
 E. Witten and especially D.J. Gross for very helpful 
conversations and discussions.
This research was supported by a Robert H. Dicke fellowship and by
DOE grant DE-FG02-91ER40671.

\end{document}